\documentclass[prd,aps,showkeys,superscriptaddress,two column]{revtex4}

\usepackage{amsmath}
\usepackage{amssymb}
\usepackage{amsthm}
\usepackage{mathrsfs}
\usepackage{graphicx}
\usepackage{fancyhdr}
\usepackage{array}
\usepackage{simplewick}
\usepackage{latexsym}
\usepackage[all]{xy}
\usepackage{eufrak}
\usepackage{euscript}
\usepackage{enumerate}
\usepackage{dsfont}
\usepackage{slashed}
\usepackage{hyperref}

\begin{document}

\title{The BF theory as an electric Julia-Toulouse condensate}

\author{L. S. Grigorio}
\affiliation{Centro Federal de Educa\c{c}\~ao Tecnol\'ogica Celso Suckow da Fonseca, 28635-000, Nova Friburgo, Brazil}

\author{M. S. Guimaraes}
\affiliation{Instituto de F\'\i sica, Universidade do Estado do Rio de Janeiro, 20550-013, Rio de Janeiro, Brazil}

\author{R. Rougemont}
\affiliation{Instituto de F\'\i sica, Universidade Federal do Rio de Janeiro, 21941-972, Rio de Janeiro, Brazil}

\author{C. Wotzasek}
\affiliation{Instituto de F\'\i sica, Universidade Federal do Rio de Janeiro, 21941-972, Rio de Janeiro, Brazil}

\author{C. A. D. Zarro}
\affiliation{Instituto de F\'\i sica, Universidade Federal do Rio de Janeiro, 21941-972, Rio de Janeiro, Brazil}

\begin{abstract}
The Julia-Toulouse mechanism is used to show that the topological Abelian BF term may be induced by the condensation of electric charges. As an application we discuss the subtle question of including consistently magnetic defects into the Maxwell-BF theory in a way to avoid the usual problems of current conservation, charge quantization, Elitzur's theorem violation and the reality of the Dirac brane, produced by the non-minimal coupling. We also discuss a new way of obtaining the Dirac's veto, which is based on the so-called Dirac brane symmetry.
\end{abstract}


\maketitle

\section{Introduction}
\label{sec:introduction}

Recently, the topological Abelian BF theory coupled to external fields has attracted attention as an effective description of topological insulators in $3D$ and $4D$ \cite{moore}. This topological theory can be seen as a long distance limit of the Maxwell-BF (MBF) theory. The MBF model generalizes for arbitrary dimensions the topological mass generation of the Maxwell-Chern-Simons (MCS) theory \cite{deser,semenoff}. In this Letter we use a generalization of the Julia-Toulouse Approach (JTA) for condensation of defects \cite{jt,qt}, put forward in \cite{artigao}, and show that the BF term can be induced through a condensation of electric charges. As an application we discuss the inclusion of monopoles in topologically massive theories without running into the contradictions that the non-minimal substitution produces.

The JTA determines the effective model describing the regime of an Abelian gauge theory with condensed defects. Julia and Toulouse \cite{jt} postulated that the condensate of defects establishes a new medium: a continuous distribution of defects. The low energy excitations of this medium represent the degrees of freedom of the condensed regime. They specified a prescription to identify these new degrees of freedom, whose signature is the potential tensor rank-jump: the degrees of freedom in the condensed phase are represented by a higher rank tensor compared to the diluted phase. Latter, the JTA was extended to relativistic $p$-form theories in \cite{qt}, leading to a clear determination of the condensed phase effective action. 

Based on \cite{jt,qt}, and also in \cite{mvf,banks} regarding the formulation of ensembles of defects, we developed a general procedure \cite{artigao} to address the condensation of defects, whose main feature is a careful treatment, compatible with the Elitzur's theorem \cite{elitzur}, of a local symmetry called as Dirac brane symmetry (which is independent of the gauge symmetry \cite{mvf}), consisting in the freedom of deforming the unphysical Dirac strings \cite{dirac} without any observable consequences. We have also developed a dual JTA, where one is able to approach the condensation of electric currents minimally coupled to the gauge field. This is the formulation to be used here.

It is possible to induce the CS term in $3D$ through a condensation of electric charges via the JTA \cite{santiago,artigao}. The condensate (of spins) gives topological mass to the photons. The inclusion of magnetic instantons is a subtle issue, since a naive substitution of the electromagnetic curvature by a non-minimal coupling with the magnetic charges violates the Elitzur's theorem, breaking the magnetic brane symmetry and spoiling the charge quantization in the electric condensed regime; furthermore, the violation of the Elitzur's theorem in this case also spoils the charge conservation when we include external electric charges \cite{mcsmon,artigao}. 

Similar issues also arise in $4D$ \cite{MouraMelo:1998uv} \cite{Ignatiev:1995qz}. In this Letter we generalize \cite{santiago,mcsmon,artigao} and show that the topological BF term may be induced by an electric condensation. Hence, it is expected that by including magnetic sources, they should become confined due to the Meissner effect generated by the condensate. Besides, we show that, contrary to the findings in the literature, including monopoles does not violates the Elitzur's theorem, charge quantization and the vortex density conservation. As a bonus a new way of obtaining the Dirac's veto is disclosed, which is based on the magnetic brane symmetry.

\section{The BF term as an electric condensate}

We shall work in Minkowski spacetime $\mathcal{M}_{d+1}$ with metric determinant $g=-1$ and $c=\hbar=1$. The partition function in the diluted regime is,
\begin{align}
Z_d[J_1]\!\! =\!\!\!\int\!\!\!\mathcal{D}A_1 \exp\left\{-i\!\!\int\!\! \frac{1}{2} dA_1  \wedge*dA_1  + eA_1\wedge *J_1\right\}
\label{eq:1}
\end{align}
where $J_1=\delta\Sigma_2$ is the electric current, the physical boundary of the world-surface of the electric Dirac string localized by the Chern-Kernel $\Sigma_2$.

Next we apply the dual JTA. To this end we add to (\ref{eq:1}), an activation term for the electric loops (which gives dynamics to the electric branes) such that it preserves the relevant symmetries ($P$, $T$, Lorentz and the local gauge and brane symmetries) and gives the dominant contribution for the dynamics of the condensate in the infrared regime \cite{artigao,mvf}:
\begin{align}
S_{activation}[J_1]&=\int\frac{(-1)^{d}}{2\Lambda^{d-1}}J_1\wedge *J_1\nonumber\\
&=\int\frac{(-1)^{d-1}}{2\Lambda^{d-1}}d*\Sigma_2\wedge *d*\Sigma_2,
\label{eq:5}
\end{align}
where $\Lambda$ is a scale associated to the electric condensate.

Introducing a sum over (the branes Poincare-dual to) $*\Sigma_2$, we obtain the partition function for the condensed regime as:
\begin{align}
& Z_c:=  \sum_{\left\{*\Sigma_2\right\}}\int\mathcal{D}A_1  \exp\left\{i\int\left[-\frac{1}{2}dA_1\right.\right.\wedge*dA_1\nonumber\\
&-eA_1\wedge d*\Sigma_2 +\left.\left. \frac{(-1)^{d-1}}{2\Lambda^{d-1}}d*\Sigma_2\wedge *d*\Sigma_2\right]\right\}.
\label{eq:6}
\end{align}
Introducing the identity $\mathds{1}=\int\mathcal{D}*P_2\delta[*P_2 - *\Sigma_2]$ we rewrite the partition function for the condensed regime as:
\begin{align}
&Z_c=\int\mathcal{D}A_1\mathcal{D}*P_2\left(\sum_{\left\{*\Sigma_2\right\}}\delta[*P_2 - *\Sigma_2]\right)
\nonumber\\
&\exp\left\{i\int\left[-\frac{1}{2}dA_1\wedge*dA_1-eA_1\wedge d*P_2 +\right.
\right.\nonumber\\
&\left.\left. + \frac{(-1)^{d-1}}{2\Lambda^{d-1}}d*P_2\wedge *d*P_2\right]\right\}.
\label{eq:7}
\end{align}

Next, we make use of the Generalized Poisson Identity (GPI) \cite{mvf,Grigorio:2009pi}:
\begin{align}
\sum_{\left\{*\Sigma_2\right\}}\!\!\delta[*P_2 - *\Sigma_2]=\!\!\!\!\sum_{\left\{*\Omega_{d-1}\right\}}\!
\!\!\exp\left\{2\pi i\!\!\int\!\!\!*\Omega_{d-1}\wedge *P_2\right\},
\label{eq:8}
\end{align}
where $*\Omega_{d-1}$ is the brane Poisson-dual to $*\Sigma_2$. The GPI works as a geometric analogue of the Fourier transform: when the brane configurations on the left hand side of (\ref{eq:8}) proliferate (condense), the brane configurations on the right hand side become diluted and vice-versa. Hence, the proliferation of the electric branes $*\Sigma_2$ is accompanied by the dilution of the branes of complementary dimension $*\Omega_{d-1}$ and vice-versa, what tells us that the branes $*\Omega_{d-1}$ must be interpreted as magnetic vortices over the condensate.

Redefining $*P_2 =: -\Lambda^{(d-1)/2}B_{d-1}$, we rewrite (\ref{eq:7}) as:
\begin{align}
Z_c= &\sum_{\left\{*\Omega_{d-1}\right\}}\int\mathcal{D}A_1\mathcal{D}B_{d-1}
\exp\left\{i\int\left[-\frac{1}{2}dA_1
\wedge*dA_1 \right.\right.\nonumber\\
&+mA_1\wedge dB_{d-1} +\frac{(-1)^{d-1}}{2}dB_{d-1}\wedge *dB_{d-1}\nonumber\\ &  \left.\left. - 2\pi\Lambda^{(d-1)/2}B_{d-1}\wedge *\Omega_{d-1}\right]\right\}.
\label{eq:9}
\end{align}
We have, therefore obtained the Maxwell-BF theory with the BF term, $mA_1\wedge dB_{d-1}$, induced directly from the JTA. Here, the topological mass $m:=e\Lambda^{(d-1)/2}$, is proportional to the condensate density $\Lambda^{(d-1)/2}$, disclosing its origin from this mechanism. The BF term determines the physical massive poles of the propagators of the fields $A_1$ and $B_{d-1}$ in the MBF theory.

We have also found a minimal coupling for the Poisson-dual current $*\Omega_{d-1}$ and $B_{d-1}$. This current is a defect over the condensate and represent regions without a complete condensation. It has the following interpretation. Under application of the exterior derivative the equation of motion for the condensate field $B_{d-1}$ implies the constraint $d*\Omega_{d-1} = 0$, which can be solved locally by the introduction of a current $*\lambda_d$ according to $*\Omega_{d-1} = d*\lambda_d$. This means that the magnetic vortices are closed and hence the magnetic vortex density is conserved: $\delta\Omega_{d-1} = 0$. Notice that in the very long distance regime, we can discard the Maxwell terms, which are of second order in the derivatives, thus obtaining from (\ref{eq:9}) the BF term, which arises here as a result of an electric condensation. This is our main result.

\section{Compatibility with magnetic monopoles}

It is known that in $3D$ the inclusion of magnetic charges via non-minimal coupling into the MCS theory spoils both charge conservation and charge quantization \cite{Henneaux:1986tt}. Besides, trying to fix such problems by adding {\it ad hoc} electric currents ends up transforming Dirac branes (an unobservable object) into real branes, therefore violating explicitly Elitzur's theorem. Also, the computation of the fermionic determinant after non-minimal coupling leads to an ill-defined object since this coupling lends the potential $A_1$ singular \cite{Fradkin:1990xy}. The incompatibility between topological mass and magnetic charges has been solved by the JTA, where a carefull treatment of the magnetic Dirac brane symmetry is possible \cite{mcsmon}. The same sort of incompatibility between topological mass and magnetic charge has also been reported in $4D$ \cite{MouraMelo:1998uv} showing that the magnetic currents can not be included in the topologically massive eletrodynamics via non-minimal coupling, leading to the same sort of difficulties. Next, we show how the JTA fixes all these problems.

The logic is simple. We introduce the magnetic charge, via non-minimal coupling, in the diluted massless phase, where it is permitted \cite{dirac}. Then we let the system condense and compute the resulting action using the JTA. Let us consider again the partition function of the Maxwell theory in the regime with diluted electric charges and external monopoles:
\begin{align}
&Z_d[J_1,j_{d-2}]=\int\mathcal{D}A_1 \exp\left\{i\int\left[-\frac{1}{2}(dA_1+\right.\right.\nonumber\\
&\left.\left. -g*\chi_{d-1})\wedge*(dA_1-g*\chi_{d-1}) - eA_1\wedge *J_1\right]\right\},
\label{eq:1b}
\end{align}
where we added to (\ref{eq:1}) the monopole current $j_{d-2}=\delta\chi_{d-1}$, the physical boundary of the world-hypersurface of the Dirac hyperstring (magnetic brane) localized by the Chern-Kernel $\chi_{d-1}$.

The non-minimal coupling $(dA_1-g*\chi_{d-1})$ describes the physical electromagnetic field in the presence of magnetic defects \cite{dirac,mvf}: the gauge field $A_1$ as well as $dA_1$ are singular over the magnetic branes and the singularity of $dA_1$ is canceled out by $*\chi_{d-1}$, lending it regular everywhere. The minimal coupling of $A_1$ with the current $*J_1$ vanishes almost everywhere, being non-trivial only along the world-line of the electric charge. If the magnetic branes touch the electric world-line at any point, the minimal coupling becomes singular, since $A_1$ is singular over the magnetic branes. Hence, for the minimal coupling to be regular along the whole trajectory of the electric charge, the magnetic branes must not touch the electric world-line: this is the \emph{Dirac's veto} \cite{dirac}.

The Dirac's veto as well as the charge quantization actually come together from the brane symmetry, corresponding to the unobservability of the unphysical Dirac branes. Under an arbitrary local deformation of the magnetic hyperstrings attached to the monopoles:
\begin{align}
*\chi_{d-1}&\mapsto *\chi_{d-1} + d*\tau_d,\nonumber\\
A_1&\mapsto A_1 + g*\tau_d,
\label{eq:2}
\end{align}
where $*\tau_d$ is a delta-distribution that localizes the hypervolume spanned by the deformation of the magnetic brane keeping fixed its physical boundary corresponding to the monopole current, the non-minimal coupling $(dA_1-g*\chi_{d-1})$ is trivially invariant, while the minimal coupling modifies (\ref{eq:1b}) by the following factor:
\begin{align}
\exp\left\{-ieg\int*\tau_d\wedge *J_1\right\}=\exp\left\{-iegN\right\},
\label{eq:3}
\end{align}
where the integer $N$ is the intersection number between (the branes Poincare-dual to) $*\tau_d$ and $*J_1$. Hence, in order to the theory to be invariant under deformations of the unphysical Dirac strings, one must impose as a consistency condition the \emph{Dirac charge quantization} \cite{dirac,mvf}: $eg = 2\pi n, n\in\mathbb{Z}$. Suppose that initially the magnetic branes do not touch the electric world-lines. The minimal coupling is then regular along the whole electric world-lines. If the deformed magnetic branes cross the electric world-lines, then at the points where these crossings happen the minimal coupling is singular and hence the operation of deforming the Dirac branes is not a symmetry of the system, since we begin with a regular minimal coupling and end up with a singular one. Hence, for the brane transformation (\ref{eq:2}) to be a symmetry of (\ref{eq:1b}), we must require that the magnetic branes never cross the electric world-lines at any point and this is the Dirac's veto. Thus, \emph{the Dirac's veto as well as the charge quantization follow as consequences of the brane symmetry} (\ref{eq:2}). We emphasize that this is a new way of deriving the Dirac's veto: the usual argument in the literature involves the analysis of the equation of motion of an electric charge in the presence of monopoles \cite{dirac} and not the much simpler analysis of the brane transformation as we did here.

A few comments are in order. First, notice that in $(d+1)$-dimensions a closed curve (the electric world-line) can cross a $d$-volume without crossing its boundary (which is the union of the original magnetic Dirac brane and the deformed one). Hence, the intersection number in (\ref{eq:3}) can be non-trivial in general and still respect the Dirac's veto, since what is involved in (\ref{eq:3}) is only the intersection number between the $d$-volume spanned by the deformation of the magnetic branes (in a space of $(d+1)$-dimensions) and the electric world-line. However, the Dirac's veto automatically excludes all kinds of deformations of the magnetic Dirac branes that implies crossing the deformed magnetic Dirac branes with the electric world-lines. In this way, \emph{the magnetic (Dirac) brane symmetry corresponds actually to the transformation (\ref{eq:2}) subjected to the constraint of the Dirac's veto, or in other words, the brane symmetry corresponds to the freedom of moving the unphysical magnetic Dirac strings through the geometric place not occupied by the electric charges}.

Next, we allow the electric charges to establish the continuum condensate. Following the previous steps, we get:
\begin{align}
&Z_c[j_{d-2}]=\sum_{\left\{*\lambda_d\right\}}\int\mathcal{D}A_1\mathcal{D}B_{d-1}
\exp\left\{i\int\left[-\frac{1}{2}(dA_1+\right.\right.\nonumber\\
&-g*L_{d-1})\wedge*(dA_1-g*L_{d-1})+mA_1\wedge dB_{d-1} +\nonumber\\
&\left.\left. +\frac{(-1)^{d-1}}{2}dB_{d-1}\wedge *dB_{d-1}\right]\right\},
\label{eq:11}
\end{align}
after using the shift $A_1\mapsto A_1 + \frac{2\pi}{e}*\lambda_d$ and the Dirac quantization condition to define the magnetic brane invariant:
\begin{align}
*L_{d-1} := *\chi_{d-1} - d*\lambda_d.
\label{eq:12}
\end{align}
This is our cherished result: we were able to obtain the action for the topologically massive theory (MBF) written only in terms of brane-invariants and completely compatible with Elitzur's theorem. This action is therefore free of all illnesses mentioned before. 

It instructive to put this result in perspective. To this end we return to the issue of the brane symmetry, but in the context of the condensed regime (\ref{eq:11}). The goal is to elucidate the issue, wrongly established in the literature, that the Dirac brane becomes real. As stated before, the brane symmetry corresponds to the freedom of moving the unphysical Dirac strings through the geometric place not occupied by the electric charges. Notice that the ``size'' of this geometric place is not defined \emph{a priori}: it depends on the electric content of the system. In the condensed regime, the electric world-lines proliferated in such a way that they established a continuum, which is described here by the condensate field, $B_{d-1}$. Hence, due to the Dirac's veto, \emph{in the condensed regime the only place allowed for the magnetic Dirac strings is the interior of the magnetic vortices connected to the monopoles}. In such a setup, which can be read off from (\ref{eq:12}) with non-trivial $*\chi_{d-1}$ (in regions where $*\chi_{d-1} = 0$ and $*\Omega_{d-1}=d*\lambda_d\neq 0$ we have from (\ref{eq:12}) the closed vortices disconnected from the monopoles), the flux inside the Dirac strings cancels out part of the flux inside the closed vortices, leaving only the open vortices with a pair of monopole-antimonopole in their ends. This is a new and important result. These (composite) open vortices are the real objects in the theory (emerging from the spontaneous breaking of the brane symmetry) and correspond to the confining magnetic flux tubes, as we show next.

To compute the confining potential for the monopoles we especialize for $D=4$ and integrate out the gauge fields $B_{2}$ and $A_1$ in (\ref{eq:11}) using Lorentz gauge conditions:
\begin{align}
Z_c[j_1]&=\exp\left\{i\int-\frac{g^2}{2}j_1\wedge \frac{1}{-\Delta+m^2}*j_1\right\}\nonumber\\
&\sum_{\left\{*L_2\right\}}\exp\left\{i\int\frac{m^2g^2}{2}L_2\wedge\frac{1}{-\Delta+m^2}
*L_2\right\},
\label{eq:16}
\end{align}
where we used that $j_1 = \delta\chi_2 = \delta L_2$ and considered a complete dilution of the closed vortices disconnected from the monopoles, such that the sum over configurations is now taken over all the possible shapes of the brane invariants \cite{artigao}.

Considering a stationary monopole-antimonopole configuration and the asymptotic time regime $T\rightarrow \infty$, the dominant contribution in the sum (\ref{eq:16}) is given by a linear magnetic flux tube corresponding to the minimal distance between the monopoles, giving the minimal energy of the system \cite{artigao}. In this limit, the second term in (\ref{eq:16}) gives a confining potential, while the first gives a short-range Yukawa interaction \cite{artigao}.

To conclude, it is important to contrast our approach versus the naive non-minimal coupling inclusion of monopoles directly into the theory's condensed regime. The partition function (\ref{eq:11}) defines the MBF theory in the presence of external monopoles consistent with the Elitzur's theorem. The naive introduction of monopoles into the MBF consisting in the substitution $dA_1\mapsto (dA_1-g*\chi_{d-1})$ in the theory (which appears in the Maxwell term, $dA_1\wedge*dA_1$, and also in the BF term, $B_{d-1}\wedge dA_1$), would not take into account the ensemble of internal defects $\left\{*\lambda_d\right\}$. In such a case, the brane symmetry (\ref{eq:2}) would be violated and the charge quantization would be spoiled in the condensed regime (basically, without taking the ensemble of internal vortices $\left\{*\lambda_d\right\}$ into account, the unphysical Dirac strings ``would become real, constituting the confining flux tubes''). Hence, it is clear that in the presence of external magnetic sources it is impossible to have a complete condensation, what would imply in the complete dilution of the ensemble of magnetic vortices $\left\{*\lambda_d\right\}$, destroying the brane invariants and spoiling the brane symmetry. Such a restriction over the condensation when there are external magnetic sources embedded in the system is easily understandable in physical terms: the Meissner effect (generated by the topological mass $m$) expels the magnetic fields generated by the external monopoles of almost the whole condensate, however, these fields cannot simply vanish - they become confined in regions with minimal volume corresponding to the magnetic confining flux tubes described by the brane invariants $*L_{d-1}$. The naive prescription for the inclusion of monopoles into the MBF system also violates the magnetic vortex density conservation, what can be seen from the equations of motion obtained from the variation of the field $B_{d-1}$ in such a naive effective action if we add a minimal coupling of the field $B_{d-1}$ with an external vortex current. This problem does not happen in our approach, as discussed after equation (\ref{eq:9}).

\section{Conclusion}

In this Letter we used the JTA to induce the topological BF term as a condensation process and discussed how to include external magnetic defects into the MBF theory without violating the Elitzur's theorem, obtaining in the condensed regime the magnetic confinement, the charge quantization and the vortex density conservation in a consistent way. The strategy was to include the magnetic charges in the diluted phase and only after that let the JTA induce the BF term that turns the theory massive. These are our mains results.

The brane symmetry is seen to imply both, charge quantization and Dirac's veto, the latter being obtained here through a new argument based on the behavior of a minimal coupling under deformations of the Dirac strings. With the use of the GPI it is clear that the currents minimally coupled do the field of the condensate in the MBF theory are naturally interpreted as magnetic vortices describing regions where the condensate has not been established.

In the condensed regime, monopoles are confined and their presence prohibits the realization of a complete condensation, since the magnetic open vortices with the monopoles in their ends cannot be undone by the Meissner effect, which can at most dilute the closed magnetic vortices. The formation of these open vortices corresponding to the confining flux tubes is consistently obtained in the formalism here presented due to a careful account of the Dirac brane symmetry and the Dirac's veto.

\section{Acknowledgements}

We thank Conselho Nacional de Desenvolvimento Cient\'ifico e Tecnol\'ogico (CNPq) for financial support.

\end{document}